\documentclass[prl,aps,twocolumn,showpacs,preprintnumbers,amsmath,amssymb]{revtex4}
\usepackage{graphicx,epsfig}
\usepackage{bm}

\def\d{{\rm d}}
\def\bea{\begin{eqnarray}}
\def\eea{\end{eqnarray}}

\begin{document}
\title{Exactly Soluble BPS Black Holes in Higher Curvature $N=2$ Supergravity}
\author{Dongsu Bak, Seok Kim, Soo-Jong Rey} 
\email{dsbak@mach.uos.ac.kr, seok@kias.re.kr, sjrey@phya.snu.ac.kr}
\affiliation{
Physics Department, University of Seoul, Seoul 130-743 KOREA\\
 School of Physics, Korea Institute for Advanced Study, Seoul 130-012 KOREA\\
 School of Physics, Seoul National University, Seoul 151-747 KOREA}
\begin{abstract}
We find a class of d=4, N=2 supergravity with $R^2$-interactions
that admits exact BPS black holes. The prepotential contains
quadratic, cubic and chiral curvature-squared terms. Black hole
geometry realizes stretched horizon, and consists of anti-de Sitter,
intermediate and outermost flat regions. Mass and entropy depends on
charges and are modified not only by higher curvature terms but also
by quadratic term in the prepotential. Consequently, even for large
charges, entropy is no longer proportional to mass-squared.
\end{abstract}
\pacs{04.70.-s 11.25.-w 04.65.+e } 
\keywords{black hole, string theory, supergravity}
\maketitle
Recently, in string theory, black holes in higher curvature gravity received renewed interest. For one thing, an interesting connection \cite{osv} was found between the partition function of Bogomolnyi-Prasad-Sommerfield (BPS) black holes in quantum Type II string theory compactified on a Calabi-Yau 3-fold and the product of partition sums for topological and anti-topological strings. It is claimed that black hole entropy, which is modified by higher curvature and higher genus effects, is interpretable as a Legendre
transform of free energies for the topological and anti-topological strings. For another, for BPS black holes with fewer charges, it was discovered \cite{dabholkar} that the stretched horizon as envisaged by Sen \cite{sen} emerges naturally by the higher curvature and
higher genus effects. See \cite{osvrecent}, \cite{horizonrecent} for subsequent works for these issues.

In $d=4, N\ge 2$ supergravity, part of higher curvature effects is encoded into so-called gravitational $F$-terms. Bosonic part of these terms takes a form of square of Riemann curvature $R^2$ times a compactification-dependent (local) function involving graviphoton field strengths and scalar fields in the vector and hyper multiplets. It was established in \cite{dewit} that universal near-horizon behavior of the black hole configuration, called attractor mechanism \cite{attrac}, remains valid even after including gravitational $F$-terms. Aforementioned works \cite{osv, dabholkar} are both built essentially on this observation: black hole entropy encodes information of gravitational $F$-terms. Despite such development, no analytic solution for BPS black holes is currently available once higher curvature terms are
incorporated. In this Letter, we report discovery of a class of exactly soluble $1\over 2$-BPS black holes in $d=4, N \ge 2$ supergravity with gravitational $F$-terms. They are built upon a simple variant of the supergravity prepotential, and hence are expected to serve as invaluable reference for better understanding of topological string, black hole entropy, stretched horizon and beyond.

We shall adopt the method of superconformal multiplet calculus in constructing $N=2$ Poincar\'e supergravity coupled with matter, and follow closely the formalism developed in \cite{dewit}. Denote the vector multiplet as $(X^I, F^{+I}_{ab})$, the hypermultiplet
potential as $\chi$, the auxiliary multiplet as $(D, T_{ab})$, the Weyl curvature-squared multiplet as $(\hat{A}, \hat{C})$, and the prepotential as $F=F(X^I, \hat{A})$ with notations $F_I \equiv (\partial F / \partial X^I)$, $F_A \equiv \partial F / \partial \hat{A}$. Then relevant bosonic part of the Lagrangian reads
\bea && 8 \pi e^{-1} {\cal L} \nonumber \\
&=& i {\cal D}_\mu F_I {\cal D}^\mu \overline{X}^I - i F_I \overline{X}^I ({1 \over 6}R - D) + {\chi \over 4} ({1 \over 3}R + D) \nonumber \\
&+& {i \over 4} F_{IJ} (F^{-I}_{ab} -{1 \over 4} \overline{X}^I \overline{T}_{ab})(F^{-Jab} - {1 \over 4} \overline{X}^J \overline{T}^{ab}) \nonumber \\ &-& {i \over 8} F_I (F^{+I}_{ab} -{1 \over 4} X^I T_{ab})T^{ab} -{i \over 32} F (T_{ab})^2 \nonumber \\
&+& {i \over 2} \hat{F}^{-ab} F_{AI} (F^{-I}_{ab} -{1 \over 4} \overline{X}^I \overline{T}_{ab}) + {i \over 2} F_A \hat{C}  + {\rm h.c.}\,. \eea

The $1\over 2$-BPS black hole is constructable as follows \cite{dewit}. Denote local $N=2$ supersymmetry parameters as $\varepsilon_A = (\varepsilon^A)^*$ where $A=1,2$ are SU(2) indices. The $1 \over 2$-BPS condition reads $e^{ i \alpha (x)}
\varepsilon_A(x) = \epsilon_{AB} \gamma_0 \varepsilon^B(x)$, where $\alpha(r)$ is a phase-factor to be determined. This puts the spacetime metric for a single-centered black hole in Tod's form:
\begin{eqnarray}
\d s^2 = -e^{2 g(r)} \d t^2 + e^{-2g(r)} (\d r^2 + r^2 \d \Omega^2_2), \label{metric}
\end{eqnarray}
sets all hypermultiplet scalars and hence the potential $\chi$ to constant, and imposes all vector multiplet scalars to obey to so-called ``generalized stabilization conditions":
\begin{eqnarray} {\rm Im} Y^I &=& \widetilde{H}^I \equiv {1\over2}\left(\widetilde{h}^I
+ {p^I \over r} \right) \nonumber \\
{\rm Im} F_I &=& H_I \equiv {1\over2}\left(h_I + {q_I \over r}\right). \end{eqnarray}
Here, $Y^I= e^{-g} X^I$ and $\Upsilon = e^{-2g} \hat{A}$ are scale and U(1) invariant variables of the vector and the Weyl curvature-squared multiplets, and $F = F(Y, \Upsilon)$ is the prepotential of degree 2, where $Y^I$ and $\Upsilon$ carry degree 1
and 2, respectively. Other equations to be solved together are
\begin{eqnarray} && \Upsilon = - 64 (\nabla_p g)^2  \label{upsilon} \\
&& {\rm Im} (Y^I \overline{F}_I) + {\chi \over 4} e^{- 2g} = - 128
e^g \nabla_p ({\rm Im} F_{\Upsilon} \nabla_p e^{-g}) \label{g} \\
&& \widetilde{H}^I \nabla_p H_I - H_I \nabla_p \widetilde{H}^I = 0, \end{eqnarray}
where we made use of rotational symmetry of the ansatz eq.(\ref{metric}).
All hypermultiplet scalars being constant, $\chi$ takes a negative constant value and sets an overall mass scale in the theory.

We shall consider a theory defined by prepotential:
\bea F (Y, \Upsilon) = - \left( C {Y^1 Y^2 Y^3 \over Y^0} + { i \tau \over 2} (Y^3)^2 + k {Y^3 \over Y^0} \Upsilon \right), \label{prepotential} \eea
and look for a BPS black hole solution. Specifically, we shall look for a solution carrying both electric and magnetic charges, but not of dyonic type. Choosing these charges along $0$ and $3$ directions in vector multiplets, we find the solution is obtained by taking an ansatz that $Y^0$ is purely real, $Y^1=Y^2=0$, and $Y^3$ is purely imaginary. In addition, we shall fix the phase and the scale in the superconformal symmetry by the gauge $\alpha = 0, g(\infty) = 0$. The generalized stabilization equations are solved by
\bea -i Y^3 &=& \widetilde{H} \equiv \Big(\widetilde{h} + {p \over r}\Big) = \widetilde{h} \Big(1 + {\widetilde{Q} \over r} \Big) \qquad (\widetilde{Q} > 0) \label{y3} \\ 
+i F_0 &=& H \equiv \Big(h + {q \over r} \Big) = h \Big(1 + {Q \over 4}\Big) \qquad (Q > 0). \nonumber \eea
Using (\ref{upsilon}), $Y^0$ is then solved from $F_0$ in terms of $H, \tilde{H}, g$:
\bea Y^0 = 8 \sqrt{k} g'(r) \sqrt{\widetilde{H} / H}. \label{y0} \eea
We also have Im$F_\Upsilon = - (\sqrt{k}/8g') \sqrt{H \widetilde{H}}$ and
\bea {\rm Im} (Y^I \overline{F}_I) &\equiv& {e^{-K} \over 2} = \tau
\widetilde{H}^2 + 16 \sqrt{k} g'(r) \sqrt{H \widetilde{H}} \nonumber \eea
Using these expressions, eq.(\ref{g}) becomes
\bea \tau \widetilde{H}^2 &+& 16 \sqrt{k} g' \sqrt{H \tilde{H}} + {\chi\over 4}  e^{-2g} \nonumber \\ &=& 16 \sqrt{k} e^g \, \nabla \cdot \left({1 \over g'} \sqrt{H \widetilde{H}} \nabla e^{-g} \right). \eea
The equation is further simplifiable to
\bea \tau \widetilde{H}^2 +{1 \over 4} \chi e^{-2g(r)} = - 16  {
\sqrt{k} \over r^2} \partial_r (r^2 \sqrt{H \widetilde{H}}). \label{eg} \eea
This equation determines the metric factor $e^{g(r)}$. Notice that derivatives of $g(r)$ all dropped out in eq.(\ref{eg}). This is because in our ansatz we took spherical symmetry and turned off $Y^1, Y^2$ completely. Later, we will revisit this issue when
considering more general solution.

Now, in the gauge $g(\infty) = 0$ adopted, the negative constant $\chi$ is determined as $\chi = - 4 \tau \tilde{h}^2$ (Notice that we have taken $\tau > 0$). The final form of the solution yields eq.(\ref{y3}, \ref{y0}) along with the metric factor
\bea e^{-2g(r)} = \Big(1+{\widetilde{Q} \over r}\Big)^2 + {Q \varepsilon \over r^2} \left[r^2 \sqrt{\Big(1+{Q \over r}\Big)\Big(1+{\tilde Q \over r}\Big)}\right]'  \nonumber\eea
where $\varepsilon = (16 \sqrt{k}/Q \tau \widetilde{h}^2) \sqrt{ h \widetilde{h}}$. The metric factor is remarkably simple and behaves similar to that of extremal charged black hole. To see this, choose $Q = \widetilde{Q} $. Then,
\bea e^{-2g(r)} = (1 + \varepsilon) \Big(1 + {Q \over r} \Big)^2 - \varepsilon. \eea
Near the horizon, $r \sim 0$, the spacetime metric becomes
\bea \d s^2 = (1 + \varepsilon) Q^2 \Big( {\d r^2 \over r^2} + \d \Omega_2^2 \Big) - {r^2 \d t^2 \over (1 + \varepsilon) Q^2}, \eea
and yields the Bertotti-Robinson geometry except rescaling by $(1 + \varepsilon)$. This near-horizon behavior is in fact universal for all values of $p, q$, equivalently, $Q, \widetilde{Q}$.

These black holes carry nontrivial central charge $Z$, defined by the charge coupled to the graviphoton:
\bea Z(r) \equiv e^{{1 \over 2} K(r)} \left(p^3 F_3(r) - q_0 Y^0(r) \right). \eea
The BPS mass is then determined by $Z$ at spatial infinity:
\bea M_{\rm BPS} = |Z|_{r=\infty} = \sqrt{2 \tau} p . \label{mass}\eea
The entropy as defined by surface integral of Noether current at the horizon yields
\bea S_{\rm BH} &=& \pi \Big[ |Z|^2 -256 {\rm Im} \partial_{\Upsilon} F(Y, \Upsilon) \Big]_{r=0} \nonumber \\ &=& \pi\Big[ ( 2 \tau p^2 + 32 \sqrt{k} \sqrt{pq}) + (32 \sqrt{k}
\sqrt{pq} )\Big]. \label{entropy}\eea
The first term is given by the area of the horizon, while the second term, which is not manifestly of geometric origin, is due to Weyl curvature-squared terms. This shows clearly that the black hole has a finite horizon area even for a single magnetic charge along $1$-component of the vector multiplet, and is set by the quadratic coupling $\tau$ in the prepotential. In $\tau \rightarrow 0$ limit, the entropy is the same as the one studied in \cite{dabholkar}. From the solution, this limit appears singular, but one can take a scaling limit that $\tau \sim \delta^2, \tilde{h} = 1/\delta, h = \delta$. In this limit, the black hole becomes massless, but the geometry and the entropy remains finite so long as $k$ is nonzero. Notice also that mass-entropy relation of these black holes now reads
\bea S_{\rm BH} = \pi M_{\rm BPS}^2 + 64 \pi \Big[ {k q \over \sqrt{2 \tau} } M_{\rm BPS}\Big]^{1/2 }, \eea
deviating from familiar one by higher curvature effects.

In obtaining the exact solution, we took $Y^1 = Y^2 = 0$. Actually, we can relax this and take more general ansatz $Y^1 = i \epsilon_1, Y^2 = i \epsilon_2$, where $\epsilon_{1,2}$ are real-valued constants. The stabilization equations are again solved similarly:
\bea 
Y^0 &=& \sqrt{ C \epsilon_1 \epsilon_2 + 64 k (g')^2} \sqrt{\widetilde{H}/H} \nonumber \\
Y^1 &=& i \epsilon_1, \,\qquad Y^2 = i \epsilon_2, \,\qquad Y^3 = i \widetilde{H} \nonumber \\
F_0 &=& - i H, \quad F_1 = C \widetilde{H} \epsilon_2 / Y^0, \quad F_2 = C \widetilde{H} \epsilon_1 / Y^0 \nonumber \\
F_3 &=& \sqrt{C \epsilon_1 \epsilon_2 + 64 k (g')^2} \sqrt{H/\widetilde{H}} + \tau \widetilde{H}. \eea
The equation determining $g(r)$ is now given by
\bea -{1 \over 4} \chi e^{-2 g} &=& \tau \widetilde{H}^2 + 4 \sqrt{H \widetilde{H}} {C \epsilon_1 \epsilon_2 \over \sqrt{C \epsilon_1 \epsilon_2 + 64 k (g')^2}} \nonumber \\
&+& 128 k {1 \over r^2} \partial_r \Big({ r^2 \sqrt{H \widetilde{H}} g' \over \sqrt{C \epsilon_1 \epsilon_2 + 64k (g')^2}}\Big). \label{e-g}\eea
Now, this is a complicated {\sl differential} equation for $g(r)$. Still, it clearly shows that the solution behaves essentially the same as the one we have found above in the region where $(g')^2 \gg C \epsilon_1 \epsilon_2$. Indeed, in the near-horizon region, it is
straightforward to check that $g'(r) \sim 1/r$ is a consistent solution. In the region where $(g')^2 \ll C \epsilon_1 \epsilon_2$, the first two terms on the right-hand side are of order ${\cal O}(r^0)$, and the equation is reduced to a 3-dimensional Liouville type equation. It is easy to see that the solution is {\sl oscillatory} (related observations were made in
\cite{horizonrecent}):
\bea g(r) \rightarrow g_0 +{1 \over 2r} (g_1 + g_2 \sin (\omega r + \varphi
) + {\cal O}(r^{-2}). \nonumber \eea
Expanding (\ref{e-g}) in $1/r$-series, we see that the oscillation frequency is set by
\bea \omega^2 = {1 \over 16 k}\left(C \epsilon_1 \epsilon_2 +{\tau \tilde{h}^2 \sqrt{C \epsilon_1 \epsilon_2}\over 4 \sqrt{h\tilde{h}}} \right) \eea
The coefficients $g_2, \varphi$ are undetermined at this order, but are
determinable at order $1/r^2$ and higher.

\begin{figure}
\includegraphics[width=2.in,height=2.2in,angle=0]{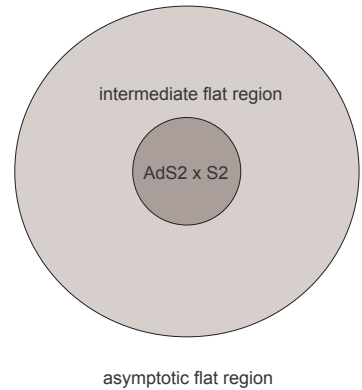}
\caption{ \label{fig1} Cartoon view of the BPS black hole.}
\end{figure}

\begin{figure}
\includegraphics[width=2.5in,height=3.2in,angle=0]{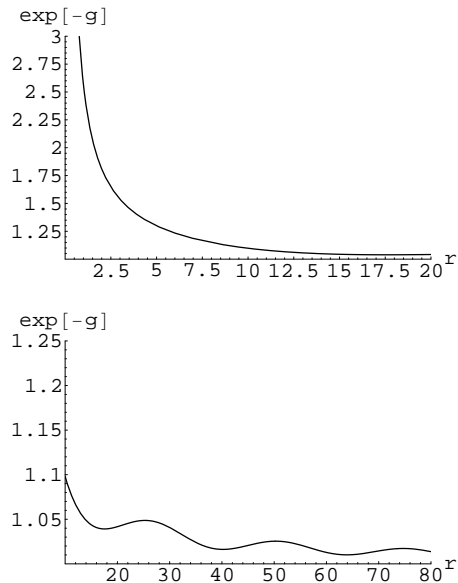}
\caption{\label{fig2} $e^{-g(r)}$ for near-horizon, intermediate, and outermost regions. We set $h = \widetilde{h}=1, p=q=1, \tau=2, 64k=1, c \epsilon_1 \epsilon_2 = 10^{-3}$, so $64k (g')^2 \sim c \epsilon_1 \epsilon_2$ around $r \sim 8$.}
\end{figure}

Black hole's global geometry thus consists of three regions: near-horizon region of $AdS_2 \times S^2$, intermediate flat region, and outermost flat region. 
See fig.\ref{fig1} for a cartoon view. 
By turning on $Y^1, Y^2$ to nonzero constant values, we have made the black hole to open up a new asymptotic region patched as the outermost region, where the metric factor exhibits oscillatory behavior. See \ref{fig2}. Actually, the oscillatory behavior can be absorbed by redefining the metric as $g_{mn} \rightarrow (1+cR) g_{mn}$.  At asymptotic infinity, by linearizing the metric as $g_{mn} = \eta_{mn} + h_{mn}$, it is straightforward to see that $h_{mn} \rightarrow h_{mn}  + cR \eta_{mn}$ for $c = -g_2/12 \omega^2$ eliminates the oscillatory part completely. Notice that it can be made effective only at the outermost region.

Behavior of the central charge $Z(r)$ is now complicated, but we will record it for reference:
\bea Z(r) = e^{K/2} \left( p^3 F_3 - q_0 Y^0 \right) \eea
where
\bea e^{-K} &=& 2 \tau \tilde{H}^2+ 4 \sqrt{H \widetilde{H}} \label{ek} \\
&\times&\Big( \sqrt{ C \epsilon_1 \epsilon_2 + 64 k (g')^2} + {C \epsilon_1 \epsilon_2 \over \sqrt{C \epsilon_1 \epsilon_2 + 64 k (g')^2}}\Big). \nonumber\eea
The BPS mass is then given again by $Z$ at spatial infinity:
\bea M_{\rm BPS} = |Z|_{r=\infty} = {\sqrt{2}\left(\tau p + \beta (p
+{\tilde{h} \over h} q) \right)\over \sqrt{\tau + 4 \beta}}, \eea
where $\beta = \sqrt{C \epsilon_1 \epsilon_2 h\tilde{h}} \slash \tilde{h}^2$. On the other hand, the entropy remains the same as (\ref{entropy}), since, near the horizon $r=0$, $\epsilon_1, \epsilon_2$ are completely negligible and the black hole behaves the same as the analytic solution found above.

Causal structure of the black hole is quite similar to Bertotti-Robinson geometry. This is most clearly seen for the special case $Q = \widetilde{Q}$. Introduce Eddington-Finkelstein coordinate $v = t + x$ where $x = r + 2Q(1 + \varepsilon) \log r - Q^2 (1 + \varepsilon)/r$. For $v=$ constant and $r \rightarrow 0$, the metric is nonsingular. Hence, the spacetime is extendible beyond the future horizon to $r<0$. We find that the
extended spacetime exhibits a timelike singularity at $r = r_\star = - (1 + \sqrt{\varepsilon / (1 + \varepsilon)})^{-1} Q$, where $g_{vv} = e^{2g(r)}$ diverges. Similar analysis and conclusion follow for the past horizon defined with coordinate $u = t - x$.

The exact black hole found above is based on the specific choice of the prepotential $F(Y, \Upsilon)$ as in (\ref{prepotential}). We shall now motivate how such a form might arise from string theory. The gravitational $F$-term, $\partial_\Upsilon^n F(Y, 0)$ originates from from one-loop $n$-point amplitudes for heterotic string theory compactified on $\mathbf{K}_3 \times \mathbf{T}^2$ or genus-$n$ amplitudes for Type II string theories compactified on a Calabi-Yau 3-fold $X$ with a structure of $\mathbf{K}_3$ fibration over $P^1(\mathbf{C})$. Each amplitude can be represented as topological partition function of topologically twisted string sigma model. It is of interest whether the prepotential (\ref{prepotential}) may be motivated from appropriate string theory compactifications. As is well-known, the special geometry of the corresponding vector multiplet moduli space ${\cal M}_{\rm V}$ is characterized by the period vector $(X^I, F_I)$ ($I=0,\cdots, n)$. The corresponding inhomogeneous special coordinates are $S = X^1/X^0$ and $T^A = X^A/X^0$ ($A=1,2,\cdots, n$). In heterotic string description, the prepotential ought to respect S-duality symmetry. Requiring covariance under the S-duality group, the prepotential becomes restricted to the form \cite{qbh}:
\bea F(X) = -{X^1 \over X^0} \eta_{AB} X^A X^B - G (X) \eea
where $G(X)$ is of degree 2 in $X^I$, invariant under the shift $S \rightarrow S + \mathbb{Z}$, and regular in the weak coupling limit $S \rightarrow + i \infty$. This implies that $G(X)$ is expandable in $m$-instanton contributions:
\bea G(X) = \sum_{m=0}^\infty e^{ 2 \pi i m S} g^{(m)}(X^0, X^A) \nonumber \eea
where again each $g^{(m)}$ is of degree 2 in $X^0, X^A$. We are interested in weak coupling limit, so set $g^{(m)}=0$ except for $m=0$. $g^{(m)}$ is also constrained to exhibit requisite singularity structure over ${\cal M}_{\rm V}$. For $g^{(0)}$, the singularity is of trilogarithmic type at gauge symmetry enhancement points. As is also well-known \cite{qbh}, T-duality symmetry imposes further constraints on $F(X)$. Such constraints are fairly complicated due to quantum corrections to the transformation laws of $X^I$'s. In particular, the dilaton $S = X^1/X^0$ is generically not invariant under T-duality. Rather, T-duality invariant dilaton $S_{\rm inv}$ takes the form
\bea S^{\rm inv} = S + {1 \over 2 (n+1)} \left( (\partial_2^2 + \cdots + \partial_n^2) g^{(0)} - {4 \over 2 \pi i} + K \right), \nonumber \eea
where $K$ is a function of degree {\sl zero}, determined up to an additive constant, and has no singularity over ${\cal M}_{\rm V}$. It thus shows that, at large $S, T^A$ far away from singular points, a given choice of $g^{(0)}$, which is of degree two, approximates specific form of quantum effects. Clearly, the simplest form is that $f^{(0)} = i \tau \delta_{AB} X^A X^B$. Its origin is somewhat trivial but is also known to be generic \cite{qbh}. By appropriate redefintion of  $X^I$'s, we thus have the structure of
$F(X)$ in (\ref{prepotential}) we started with. As for the gravitational part of $F(X)$, it typically takes the form
\bea F_{\rm grav} =  {\Upsilon} \sum_{A=1}^n{ \beta_A \over 4 \pi i}
\log \Delta(e^{4 \pi i Y^A/Y^0}), \label{gravpart} \eea
where $\Delta(q) = \eta(q)^{24}$ and $\eta(q)$ is the Dedekind eta function. for a given string compactification, threshold corrections gives rise these terms and $\beta_A (A=1,2,3)$ define gravitational $\beta$-functions. Again, at large $S, T^A$, eq.(\ref{gravpart}) is approximated by terms proportional to $Y^A/Y^0$. Our choice in (\ref{prepotential}) amounts to $\beta_1 = \beta_2 = 0$ and $\beta_3 = k$, but one can relaxed it and still find exact black holes.

We thank Gungwon Kang and Ho-Ung Yee for useful discussions. DB was
supported by KOSEF ABRL R14-2003-012-01002-0 and KOSEF R01-2003-000-10319-0. SJR was supported in part by KOSEF Leading Scientist Grant and by W.F. Bessel Award of Alexander von Humboldt Foundation.

\end{document}